\documentclass[12pt]{article}

\usepackage{epsfig}
\usepackage{amsmath}
\usepackage{soul,color}
\usepackage{bm}
\usepackage{epsfig}
\usepackage{natbib}
\usepackage{hyperref}
\newcommand{\tr}{^{\prime}}
\def\b#1{\mbox{\boldmath $#1$}}    

\renewcommand{\th}{\theta}

\newcommand{\al}{\alpha}
\newcommand{\be}{\beta}
\newcommand{\de}{\delta}

\newcommand{\si}{\sigma}
\newcommand{\ga}{\gamma}

\def\bbeta{\mbox{\boldmath$\beta$}}

\def\boeta{\mbox{\boldmath$\eta$}}

\def\bnu{\mbox{\boldmath$\nu$}}

\def\baselinestretch{1.7}
\usepackage{bm}
\usepackage{epsfig}

\begin{document}
\title{\vspace*{-1.5cm}
A causal analysis of mother's education on birth inequalities}
\author{Silvia Bacci\footnote{Department of Economics, Finance and Statistics,
University of Perugia, Via A. Pascoli, 20, 06123 Perugia.}
\footnote{{\em email}: silvia.bacci@stat.unipg.it} ,
Francesco Bartolucci$^*$\footnote{{\em email}: bart@stat.unipg.it} ,
Luca Pieroni$^*$\footnote{{\em email}: lpieroni@unipg.it}}  \maketitle
%
\def\baselinestretch{1.3}

\begin{abstract}\noindent
We propose a causal analysis of the mother's educational level on the health
status of the newborn, in terms of gestational weeks and weight. The analysis is
based on a finite mixture structural equation model, the parameters of which
have a causal interpretation. The model is applied to a dataset of almost ten thousand
deliveries collected in an Italian region. 
The analysis confirms that standard regression overestimates the impact of
education on the child health. With respect to the current economic literature,
our findings indicate that only high education has positive  
consequences on child health, implying that policy efforts in education  
should have benefits for welfare.


\end{abstract}

\newpage
\section{Introduction}\label{sec:intro}


Maternal education is known to drive wedges 
between infants' health at birth. These wedges, generally measured in terms
of weight at birth, are 
powerful determinants of health and economic outcomes as adult. A low education may yield effects 
on the initial endowment of an infant's health human capital that tends to be pervasive over the 
life. In particular, economists have argued that inequality at birth may partly be transmitted from a 
generation to the next, with the effect of a lower educational attainment, poorer 
health status, and reduced earning in adult age. 
Instead of being used as an input, birth outcomes have 
been proved to be directly affected by maternal education. Classical references that analyze the 
impact of maternal characteristics and behaviors on infant well-being   are \cite{rosenzweig:83}, 
\cite{rosenzweig:91}, and \cite{currie:03}. 

In this paper, we follow this literature and we propose a model that allows investigating whether the 
higher education can improve health child quality.
We complement the studies of \cite{almond:11}, 
\cite{case:10}, and many others, which focus on approaches for the
evaluation of the correlation between education and infant's health. The common starting 
point for our treatment is the hypothesis that this strong correlation  may partially reflect the influence of unobserved heterogeneity, 
especially when the birthweight is taken as a proxy of the health outcome.
This calls into question the existence 
of important pathways for the effect on health of education. For example, \cite{behrman:02}
find that parents with favorable heritable endowments obtain more schooling for 
themselves, are more likely to marry each other and this increases health of children. 
In addition, one would expect, for instance, that educated mothers are more likely to adopt 
behaviors 
(not smoking or drinking, enrich the 
nutritional intake, etc.) that could have a positive impact on birth outcomes. 

Different empirical strategies have been proposed in 
order to help mitigate problems of endogeneity. In an effort to isolate the 
causal effect of education on birth outcomes, one of the proposed empirical strategies uses an 
instrumental variables approach. 
Quasi-experimental infant health research, focused on primary school construction programs
\citep{breierova:04,chou:10} and on college openings \citep{currie:03}, finds the existence of a
causal effect, although the observational comparisons may even underestimate the true effect.
Unfortunately, not always from public data we are able to identify
an exogenous shock of interest or valid instruments which affects education and that may be applied within
an instrumental variable setting. 
Another approach is based on  
panel data that identifies the outcome effects at birth from changes in prenatal 
behavior or maternal characteristics between pregnancies \citep{rosenzweig:91, currie:03, abrevaya:08}. 
However, a concern about this identification strategy is the presence of feedback 
effects, specifically those of prenatal care in later pregnancies, which may be correlated with 
education and birth outcomes in earlier pregnancies.

Our paper focuses not exclusively on the key issue of how maternal education causes health 
outcomes, but it also intends  to simultaneously investigate other socio-economic relationships, such as 
the marital status on birth outcomes, exploiting the growing availability of cross-sectional
administrative data. Several approaches to causal inference have been proposed in 
the literature. Among the most known, \cite{ney:23} and \cite{rub:74} provide a definition of causal 
effects in terms of potential outcomes. We relate to
graphical models \citep{lau:96} and 
graphs of influence \citep{daw:02}, which represent extensions of path analysis \citep{wright:21}. 
Here, we heavily refer to the Pearl's approach \citep{pearl:98, pearl:00, pearl:2009, pearl:2011} 
based on the Structural Equation Models 
\citep[SEMs;][]{wright:21, gold:72, duncan:75,bol:rabe:skr:08}. Pearl extends the role of structural 
models in the econometric literature to take 
into account the causal interpretation of the coefficients, so that these models represent
mathematically equivalent alternative to the potential outcome framework 
\citep{pearl:2011}. Note that the Pearl's approach also represents an important contact point with 
other approaches. In this regard, \cite{hol:86} outlines how the path diagrams, that are commonly 
used in the SEMs, may be also used with the Neyman-Rubin's model. In addition, \cite{lau:96} and 
\cite{daw:02} place the work of Pearl in the context of graphical models and influence diagrams, 
respectively. Within the SEM causal framework, we also evaluate whether 
the educational level has effect on women's probability of being married and at which extend it can 
affect gestational age or birthweight.

In the following, we base our analysis on a SEM approach that accounts for unobserved
heterogeneity by introducing a latent background variable. 
More precisely, we propose to apply a latent class approach \citep{laza:50, 
laza:henr:68, good:74}, where we assume the existence of unobserved groups of individuals that are 
homogeneous with respect to both the cause (i.e., education and marital status) and the 
effect of interest. The resulting model is a special case of finite mixture SEM 
\citep{jed:jag:des:1997, dol:maas:98, arm:stein:witt:99, ver:mag:05}, based on a suitable number of 
consecutive equations in which: 
(\emph{i}) unobserved heterogeneity is represented by a discrete latent variable
defining latent classes of individuals, (\emph{ii}) the 
causes may depend on the discrete latent variable and on other covariates, and (\emph{iii}) 
the response variables of interest depend on the causes, on the discrete latent 
variables, and on other covariates. In this way, since the causal effect is evaluated 
within homogenous groups of individuals, it is still possible to read the partial regression 
coefficients in terms of causal effects, as it happens when we adjust for observed confounders \citep
{cox:wer:04}.

The model is estimated by an Expectation-Maximization (EM)
algorithm \citep{demp:lair:rubi:77} which is implemented by the authors through a series of {\tt R}
functions, which are available to the reader upon request.\newpage

Our contribution mainly provides new evidence on the estimation
of the causal effect of maternal education 
on infant health outcomes, attempting to control for confounding factors. We use the population of 
singleton newborns 
in a region of Italy (Umbria) from 2007 to 2009 to examine these differences in the 
usual outcomes of birthweight and gestational age. The estimation strategy controlling for
unobserved characteristics of the mother guarantees that early health outcomes are entirely 
driven by differences in education and/or marital status. 
Although the regional sample should have a smaller variability in the key variables, we shed light on 
the previous research question, highlighting that more educated mothers give birth to 
children with higher weight,
whereas gestational age is not affected by education. More strikingly, we show that 
these results unequivocally arise for mothers that have, at least, 
an educational degree when unobservable variables are taken into account by 
the introduction of latent classes. Secondly, whereas 
high school and academic qualification have a positive effect on 
the mother's probability to be 
married, this family characteristic does not appear to be a significant determinant to explain the 
inequality in birth outcomes, suggesting that these are not important for the 
effect on health. 

The outline of the paper is as follows. The next section describes the theoretical and empirical 
background. Section 3 illustrates
the data and provides preliminary evidences. Section 4 discusses our framework for quantifying the 
causal effect of maternal social characteristics on health outcomes at birth and our estimation 
strategy. In this section, we first illustrate the adopted SEM approach
and, then, we describe the estimation 
algorithm implemented to maximize the log-likelihood. Section 5 presents our main results, whose 
implications for policy-makers are discussed in Section 6. 
\newpage
\section{Conceptual framework}\label{sec:2}
The actual 
benefits of any public health initiative aimed at reducing inequality at birth 
crucially depend 
upon the estimates of the causal effect of mothers' characteristics and behaviors and the possibility 
of intervention by policy-makers. This section starts analyzing the hypothesis underlying the well-documented cross-sectional association of education and birth outcomes, such as gestational age and 
birthweight. Then, it discusses some insights linked with intermediate and confounding variables 
focusing on transmission bias of unobserved variability across mothers. 

In the following, by $\b y_{i} = (y_{i1}, y_{i2})$ we denote the vector of birth outcomes 
(gestational age, birthweight) for each singleton deliver $i$, $i=1, \ldots, n$,
by $\b z_{i}  = (z_{i1}, z_{i2})$ we denote the vector of putative causes (mother education, 
marital status), and by $\b x_{i}$ a vector of 
mother-specific characteristics (citizenship, age) other than those included in  $\b z_{i}$. 
These characteristics 
may be associated with $\b y_{i}$, but cannot be interpreted in a ``causal'' sense, being not 
modifiable in principle. Furthermore, we introduce a vector $\b u_{i}$ 
which reflects mother-specific unobservable 
determinants of child outcomes (e.g., genetic factors, unreported life style behaviors).  

Focusing only on observable variables, a simple multiple linear regression (i.e., 
through the Ordinary Least Square, OLS, estimation method) of $\b y_{i}$ on $\b z_{i}$ and $\b x_{i}$
has the regression coefficients referred to the causes in $\b z_{i}$  as central parameters 
of interest for policy purposes. Thus, if significant, 
it suggests that mothers may improve the well-being of their children with consequences on adult 
outcomes through interventions that will stimulate to attend school for a longer period of time. 
\cite{currie:11} reviews the works that study the long-term consequences of insufficient health at 
birth, confirming a strong negative association with future performance in terms of schooling 
attainment, test scores residence in high income areas, and wages\footnote{The economic literature 
has also shown the existence of selection in mothers subjected to strong deprivation with 
underestimated long-term causal impact on adults results. See the discussion of \cite{currie:11} 
about the selection effects of the second world war in German mothers. See \cite{conti:10}
for a recent discussion of women selection into higher education.}. However, ignoring the existence 
of a potentially significant vector of omitted variables, $\b u_{i}$, that simultaneously may affect 
child outcomes and putative causes, the true effect of $\b z_{i}$ results confounded, being the 
corresponding regression coefficients  under- or over-estimated. One would expect, for instance, that 
more educated mothers are more likely to adopt other behaviors, such as 
stopping smoking or drinking, that could have a positive impact on the child's characteristics.
It means that the 
OLS estimator of regression coefficients is biased because of the correlation between mother's education 
and these unobserved variables. 

With respect to the economic literature discussed in the previous section, 
we provide a different strategy of identification and estimation, following the  
SEM approach. 
We use a number of recursive concatenated equations for constructing a prediction of
birth outcomes given education. Here, we anticipate some features that make our approach suitable for the empirical test
using cross-sectional administrative data.

First, the evidence that the causes of prematurity are less well-understood with respect to
those of low birthweight does not imply that the significant 
strong correlation between the infant health outcomes 
should be obscured. Unlike \cite{almond:05}, we do not propose to model the birthweight as 
``caused" by duration of gestation because if, on the one hand, it explains large part of the overall 
variance, on the other hand, reverse causality may emerge given that lower infant weight during 
gestation may be a cause for preterm delivery. 
Therefore, we model an associative rather than a causal relationship between birthweight and 
gestational age.
 
Differently from the conventional empirical specifications, the estimates of the effects of determinants on infant 
health outcomes are not carried out by establishing thresholds of premature birth (e.g., gestation 
less than 37 weeks) and low birthweight (e.g., birthweight of infants less than 2500 grams). Indeed, 
the related economic costs have been found non-linearly significant across the distribution of 
outcomes with peaks at very preterm delivery and low end of the distribution. Following 
the aim of the paper, it is appropriate to use continuous response variables to evaluate
infant inequalities.

Second, as an important channel through which education may affect infant health outcomes, we also 
examine whether education affects the condition of a mother to be married. There are relatively few 
available theoretical models of marriage markets with pre-marital investment in education. A recent 
work in this topic is given by \cite{Chiappori:09}, which focuses on 
the role of background characteristics on the probability to be matched with skilled partner. 
On the other hand, a strand of the empirical literature focuses on the effect of 
assortative mating\footnote{The literature on assortative mating addresses
the question of who marries whom, as well as 
who marries and who remains single \citep{Becker:81}.
Positive assortative mating on a certain characteristic means that 
individuals tend to match with partners who are similar with respect to that characteristic.}, 
finding a positive impact by education \citep{pencavel:98, qian:98}. 
Here, we directly model 
the probability to be married of mothers with different 
educational level, 
leaving out heterogeneity that concerns assortative mating. Indeed,  this relationship has 
itself an interpretation within the market marriage theory. On the other hand, 
studies on assortative mating based on own education as well as background characteristics may be 
difficult to interpret, as the 
education distributions for men and women in the population differ.
As a result, the educational level should increase the probability that 
the mother is married at the 
time of the birth because at least the \textit{production of household public goods} determines gains 
from marriage as, for example, a "high quality"
child \citep{Becker:85}. 
With respect to this literature, 
the potential gains from the household specialization have been recently reduced in the developed countries, 
given a growing participation of women in the labor market and marriage postponement of more educated people.
The disencentive to marry has gone hand-in-hand with an increase in cohabitation, which work as much as marriage.

Third, a major difficulty in many specific observational studies concerns whether all appropriate 
background
variables have been included in the model to ensure that the relevant regression coefficients capture 
the causal effect
of $\b z_i$ on $\b y_i$, so that the term "cause" is appropriate for $\b z_i$. For example, 
it is known that mothers of some ethnic groups could be predisposed, \emph{coeteris paribus}, to 
give birth to children with higher birthweight. 
Since this large weight is statistically independent from education, this implies that causal 
estimation of the effect of
education on well-being of infant should control for factors that directly or indirectly 
cannot be influenced by policy interventions. 
It is known that the main limitation of many observational studies is the possible existence of unobserved 
confounders whose omission seriously distorts the interpretation of the dependence of interest, also 
when we control for many observed variables. To solve this problem, 
we extend the SEM approach to a framework with latent classes potentially able to account for 
unobserved heterogeneity. 

Fourth, in the proposed approach intermediate variables are marginalized out, 
leading to their exclusion from the analysis. For instance, intra-uterine growth 
retardation (IUGR)\footnote{See \cite{Kramer:87} 
or \cite{almond:05} for a deepen discussion in using IUGR as a proxy of inequality in health child.}
is a measure of inequality during the pregnancy, that anticipates infant outcome at birth and 
is a serious candidate for being an intermediate variable. Indeed, as a main source of variation of 
birthweight, including IUGR is irrelevant to detect the potential \emph{total} causal impact of 
mother's characteristics on infant health outcomes.
We may also include, in this set of intermediate results of education that may not be valid for 
predicting the investigated causal effect, if mothers smoke or not, a characteristic found to 
adversely impact on birthweight\footnote{See \cite{currie:03, almond:05, abrevaya:08}.}, along with 
prenatal care visits and parity, as discussed in \cite{currie:11}.

To summarize, in our theoretical model 
we assume that age and citizenship (both of them included in the vector $\b x_i$) are attributes of 
women that are not modifiable,  educational level ($z_{i1}$)  may have a causal effect on  marital 
status ($z_{i2}$) and both educational level  and marital status may have a causal effect on 
gestational age ($y_{i1}$) and birthweight ($y_{i2}$). Moreover, we assume that  gestational age and  
birthweight are inequality indicators with a likely high level of association, but without a 
specific causal relationship. 

In the following, after describing the dataset used for our analysis,
we will return to the discussion of the 
proposed approach with more technical details.
\section{Data and preliminary analyses}\label{sec:data}
In this section we first describe the dataset used for our study,
presenting some preliminary analyses that support significant 
correlations between the key covariates, 
such as education and marital status, and outcomes at birth. 
\subsection{The dataset}  \label{sec:data}
The study is based on data obtained from the Standard Certificates of Live Birth (SCLB) collected in 
Umbria (Italy) in 2007, 2008, and 2009. The SCLB is filled in within ten days after the delivery by 
one of the attendants the birth (e.g., doctor, midwife) and it provides information on infants' and 
mothers' characteristics. 
These data concern socio-economic and demographic characterisic of each 
mother giving a newborn including maternal age, citizenship, educational attainment, marital status, 
childbearing history, prenatal care, and geographic residence, whereas, as anticipated 
in Section \ref{sec:2}, the dataset 
does not contain the smoking habits before and during the pregnancy. The information on the father is 
more sparse and includes sex, citizenship, and education. Linked with the newborn, information 
include gestational age, birthweight, and pluriparity.
The available dataset contains information about over than 25,000 women. For our study we limited our 
attention to natural conceptions (i.e.,
without assisted fertilization methods), primiparous women, 
and singleton births; moreover, only infants with a gestational age of at least 23 weeks and a 
birthweight of at least 500 grams are taken into account. The total sample size that merges each 
mother and her baby amounts to 9,005 records.

Definitions and some descriptive statistics for the variables used in the successive analysis are 
shown in Table \ref{Tab1}. As already mentioned, the main attention is focused on the possible effect 
of maternal social characteristics on the inequalities in gestational age and in birthweight  of 
infants. Both variables result normally distributed (Figure \ref{fig:hist})\footnote{These results are also in line with kernel density estimates.}; 
in average, the delivery takes place after 39.310 
weeks of gestation (standard deviation equal to 1.686) 
and the mean weight at birth is equal to 3.262 kg (with a standard deviation equal to 0.487 kg). The 
two variables present a positive and intermediate correlation ($\rho = 0.56$), as confirmed by the 
scatter plot in Figure \ref{fig:graph1}.

Commenting Table \ref{Tab1}, we observe that women are on average 30 years old; the $80\%$ is 
Italian, whereas the $13\%$ comes from East-Europa. In addition, more than one half ($52\%$) has a high
school diploma, followed by a $28\%$ with a higher 
educational level (degree or above); the remaining $20\%$ 
of women attained at most a compulsory educational level. 
The $70\%$ of women is married. 
As a limitation, our 
dataset does not contain information about cohabitation but includes it on the category "not married"; this may imply a bias of
the effect of marital status on child's outcomes, under-estimating the positive effects of the production of public goods in the household. 

\begin{table}[!ht]
\begin{center}
\begin{tabular}{llrrr}
  \hline
Variable	&	Category	&	\%	&	Mean	&	St.Dev. \\ 
  \hline
Gestational age (weeks)	&		&		&	39.310	&	1.686 \\
Birthweight (kg)	&		&		&	3.262	&	0.487 \\
\hline
Age (years)	&		&		&	30.040	&	5.288 \\
\hline
Citizenship	&	Italian	&	80.1 	&		&	 \\
	&	east-Europe	&	12.6	&		&	 \\
	&	other citizenship	&	\;7.3 	&		&	 \\
\hline
Education level	&	middle school or less	&	19.8	&		&	 \\
	&	high school	&	51.9	&		&	\\
	&	degree and above	&	28.4	&		&	\\
\hline
Marital	status &	married	&	70.0	&		&	\\
	&	not married	&	30.0	&		&	\\
  \hline
\end{tabular}
\end{center}
\caption{\em Distribution of variables}
\label{Tab1}
\end{table}

\begin{figure}[!ht]\centering
\includegraphics[width=8cm, height=7cm]{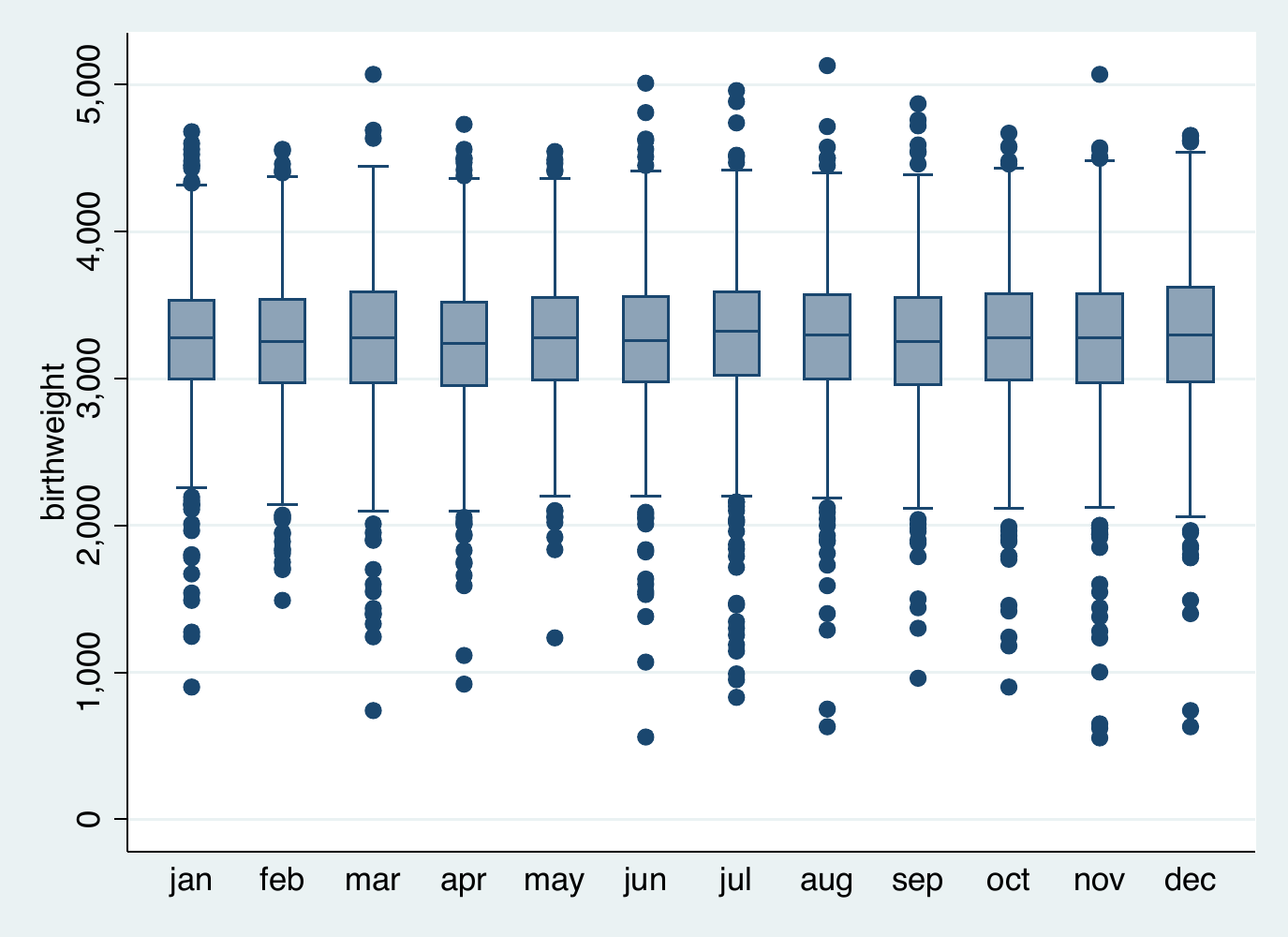}
\caption{\em Box plots for birthweight by birth month}
\label{fig:graph2}\vspace*{0.5cm}
\end{figure}

\begin{figure}[!ht]\centering
\begin{tabular}{cc}
 \includegraphics[width=8cm, height=7cm]{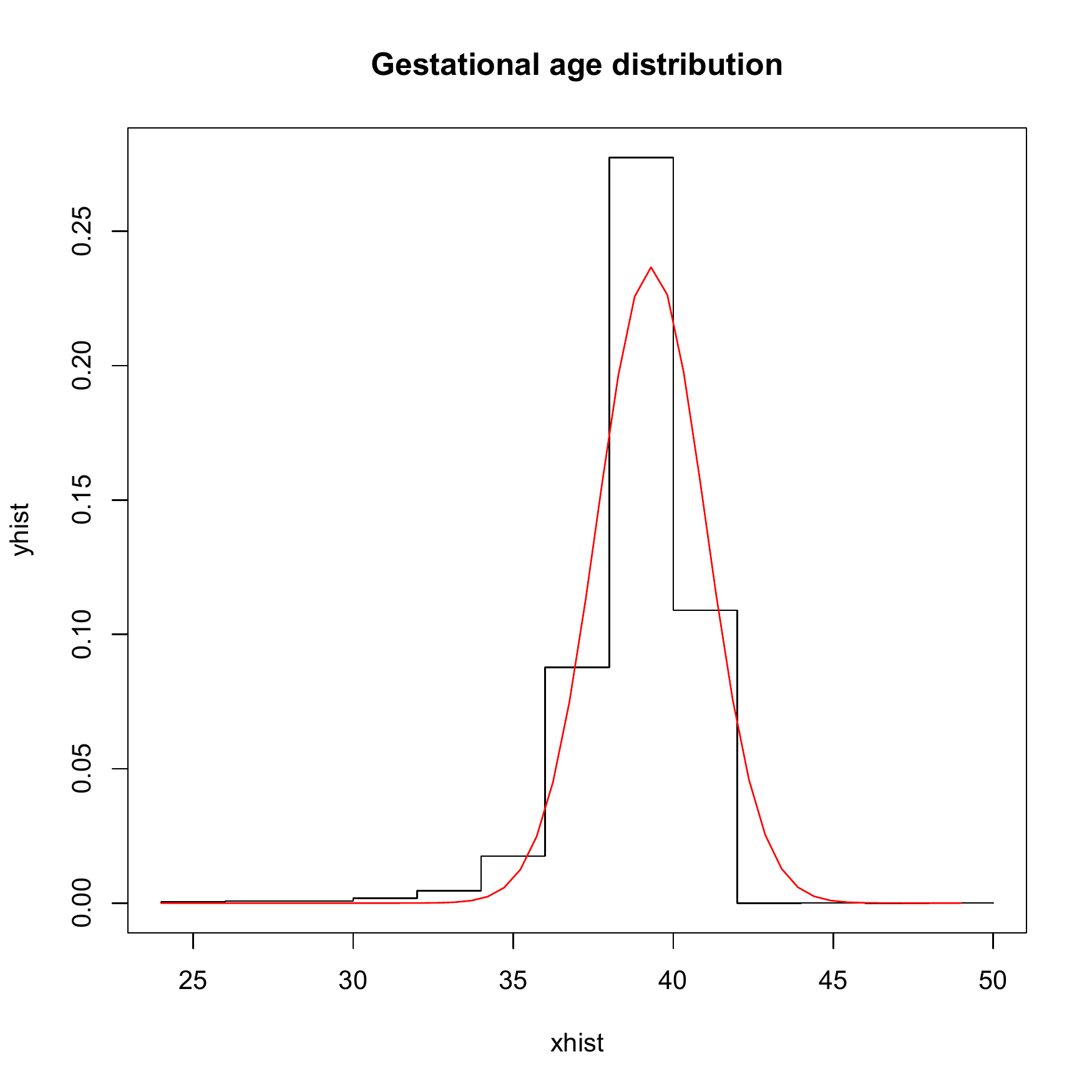} &
\includegraphics[width=8cm, height=7cm]{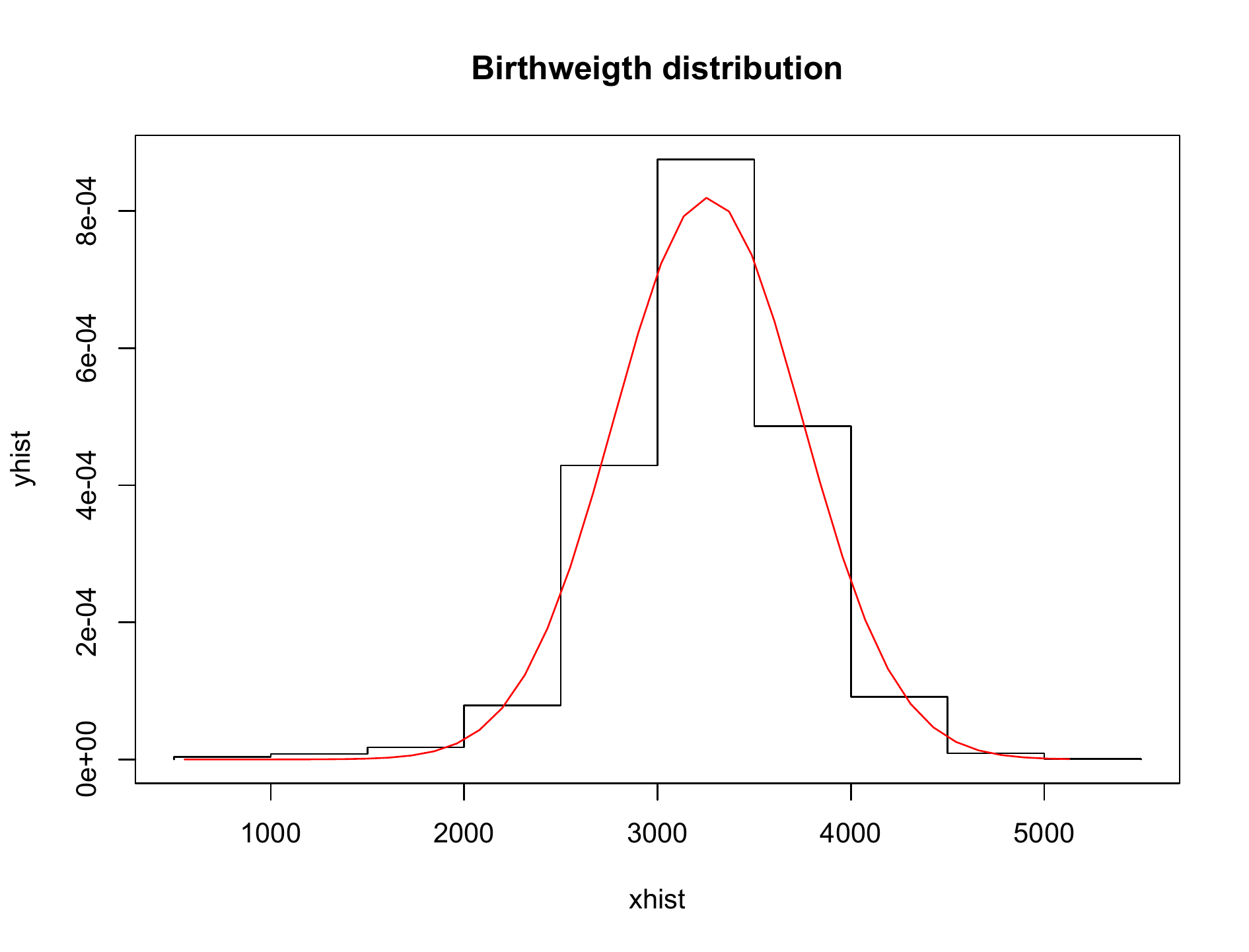}  \\
\end{tabular}
\caption{\em Histogram and normal curve for gestational age (left) and birthweight (right)}
\label{fig:hist}
\end{figure}

\begin{figure}[!ht]\centering
\includegraphics[width=8cm, height=7cm]{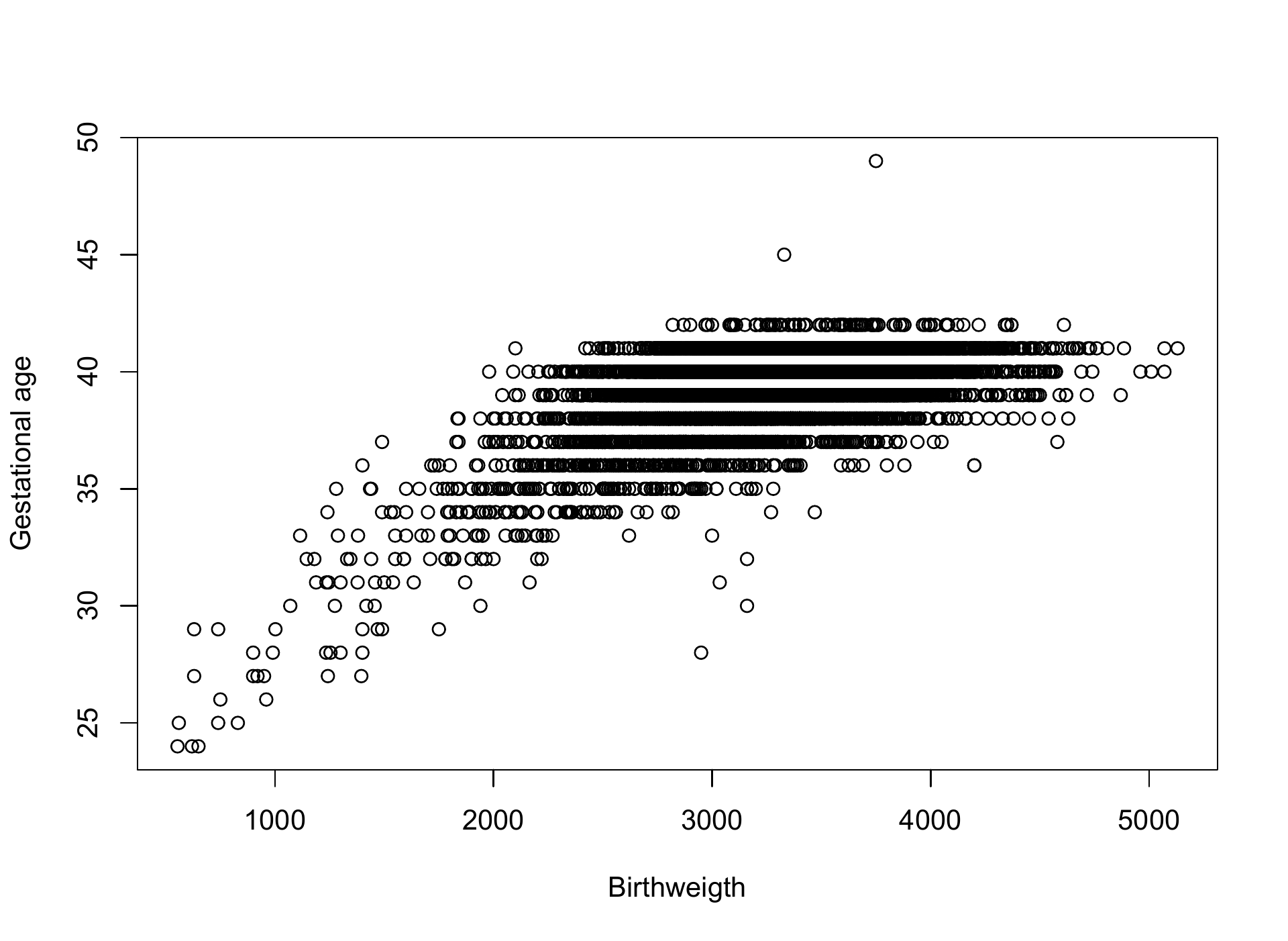}
\caption{\em Scatter plot: birthweight versus gestational age}
\label{fig:graph1}\vspace*{0.5cm}
\end{figure}

Lastly, in order to account for other potential effects on infants' weight, we report 
a box-plot of birthweight by birth month. 
The graph in Figure \ref{fig:graph2}
shows a small variability of birthweight over the year  and it corroborates the results obtained by an Anova test, allowing to strongly reject the hypothesis that winter babies tend to be smaller than the ones born in summer months, discussed in the literature by
\cite{Torche:10}. 

\subsection{Multiple regressions} \label{sec:prelim}
Through multiple linear regressions performed separately for \emph{gestational age} and 
\emph{birthweight},  it is possible to discover some significant associations with the covariates 
of Table \ref{Tab1}. As shown in Tables \ref{tab:reg_gest} and \ref{tab:reg_weight} respectively,
the number of gestational weeks and the weight of the infant
tend to decrease as the woman age increases. As concerns the 
gestational age, we observe a significant association with citizenship, with 
women coming from foreign countries tending to deliver before Italian women.
No other significant effect emerges with for the other covariates. 

Some differences exist in the associative relations between birthweight and 
the covariates. The statistical effect of citizenship is significant at $10\%$ level for 
east-European women and at $5\%$ level for other citizenships (with respect to Italian
citizenship); the birthweight results greater 
for east-European women with respect to Italians and smaller for the other citizenships. We also 
observe a significant association between birthweight and educational level: as 
expected, the higher the educational level, the greater the birthweight. Finally, some evidence of 
association is observable for the marital status: not married women give birth to 
infants with a lower weight with respect to married women.

\begin{table}[!ht] 
\begin{center}
\begin{tabular}{llrrrr}
  \hline
  \multicolumn1c{covariate} & \multicolumn1c{category}
         & \multicolumn1c{est.} & \multicolumn1c{s.e.}  & \multicolumn1c{$t$ stat.}   & \multicolumn1c{$p$-value} \\ 
  \hline
intercept & \multicolumn1c{--} & 39.325 & 0.051 & 772.686 & 0.000 \\ 
  age &  \multicolumn1c{--} & -0.019 & 0.004 & -4.910 & 0.000 \\ 
  age$^2$ & \multicolumn1c{--} & -0.001 & 0.001 & -1.336 & 0.181 \\ 
    citizenship & Italian & 0.000 & \multicolumn1c{--} & \multicolumn1c{--} & \multicolumn1c{--} \\ 
  citizenship & east-Europa & -0.242 & 0.059 & -4.099 & 0.000 \\ 
  citizenship  &  other citizenship & -0.208 & 0.072 & -2.887 & 0.004 \\ 
     education & middle school or less & 0.000 & \multicolumn1c{--} & \multicolumn1c{--} & \multicolumn1c{--} \\
  education & high school & 0.077 & 0.049 & 1.551 & 0.121 \\ 
  education & degree or above & 0.077 & 0.057 & 1.345 & 0.179 \\ 
     marital & married & 0.000 & \multicolumn1c{--} & \multicolumn1c{--} & \multicolumn1c{--} \\
  marital & not married  & -0.025 & 0.039 & -0.640 & 0.522 \\ 
   \hline
\end{tabular}
\caption{\em Regression results for the gestational age} 
\label{tab:reg_gest}
\end{center}
\end{table}

\begin{table}[!ht] 
\begin{center}
\begin{tabular}{llrrrr}
  \hline
 \multicolumn1c{covariate} & \multicolumn1c{category}       & \multicolumn1c{est.} & \multicolumn1c{s.e.}  & \multicolumn1c{$t$ stat.}   & \multicolumn1c{$p$-value} \\ 
  \hline
intercept & \multicolumn1c{--} & 3.240 & 0.015 & 220.413 & 0.000 \\ 
  age & \multicolumn1c{--} & -0.005 & 0.001 & -4.159 & 0.000 \\ 
  age$^2$ & \multicolumn1c{--} & -0.000 & 0.000 & -0.875 & 0.381 \\ 
      citizenship & Italian & 0.000 & \multicolumn1c{--} & \multicolumn1c{--} & \multicolumn1c{--} \\ 
  citizenship & east-Europa & 0.032 & 0.017 & 1.847 & 0.065 \\ 
  citizenship & other citizenship & -0.050 & 0.021 & -2.414 & 0.016 \\ 
    education & middle school or less & 0.000 & \multicolumn1c{--} & \multicolumn1c{--} & \multicolumn1c{--} \\
  education & high school & 0.032 & 0.014 & 2.243 & 0.025 \\ 
  education & degree or above & 0.050 & 0.017 & 3.033 & 0.002 \\ 
       marital & married & 0.000 & \multicolumn1c{--} & \multicolumn1c{--} & \multicolumn1c{--} \\
  marital & not married & -0.019 & 0.011 & -1.682 & 0.092 \\ 
   \hline
\end{tabular}
\end{center}
\caption{\em Regression for the birthweight}
\label{tab:reg_weight}
\end{table}

The results obtained by the multiple linear regressions suggest to proceed in the analysis 
investigating the causal relationships between variables. For this aim, we adopt a SEM framework
that will be illustrated in the next section.
\section{Proposed approach} \label{sec:model}
In the following, we resume the conceptual framework described 
in Section \ref{sec:2} and we describe in more 
detail the resulting statistical model which is used to analyze 
possible effects of social characteristics of women on 
the inequalities in gestational age and birthweight of infants. We first present some preliminary 
concepts about causality. Secondly, we describe the adopted SEM, 
which is distinguished with respect to 
standard SEMs for two main elements: 
(i) it is based on the presence of a discrete, rather than continuous, latent 
variable and (ii) it accommodates for any type of responses rather than only for continuous 
responses. After having described the structural equations of the proposed model, 
we illustrate maximum likelihood estimation of their parameters. 
\subsection{Preliminaries on causality}
\begin{figure}[h!]\centering
\includegraphics [width=13cm, height=5cm] {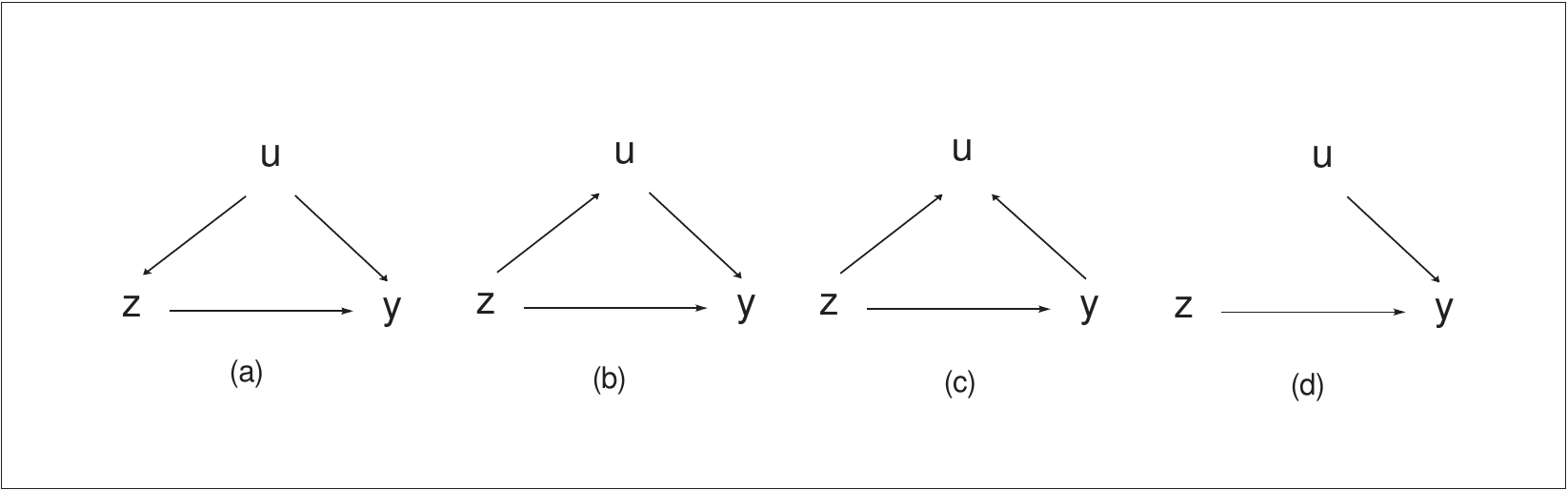}
\caption{\em Causal relation between $z$ and $y$ and presence of a third variable $u$: (a) $u$ as common cause, (b) $u$ as intermediate effect, (c) $u$ as common effect, (d) $u$ as cause acting independently from $z$}\vspace*{0.5cm}
\label{Fig1}
\end{figure}

One of the main reasons of systematic bias, which may lead to a wrong causal analysis and 
is commonly encountered in observational 
studies, is the presence of confounding effects that are ignored during the analysis \citep{Her:Rob}. 
We speak about 
confounding effect \citep{Freed:99} when two variables $z$  and $y$  have a common cause $u$, 
that confounds the true relationship between the putative cause $z$ and the effect $y$ 
(Figure \ref{Fig1} panel (\emph{a})). A typical situation  takes place when a strong observed 
association between variables $z$ and $y$ may be explained partly or completely by controlling for 
the common cause $u$. On the other hand, we can also encounter 
the opposite situation when the 
true causal relationship between $z$ and $y$  is balanced and cancelled (so that $z$ and $y$ result 
statistically independent) by a relationship of equal strength but opposite sign due to the common 
cause $u$. In all these cases, it is important to take explicitly into account all the possible 
sources of heterogeneity of $y$, in order to avoid confounding effects. 

Before proceeding, it is useful to point out that the presence of a common cause of both $z$ and $y$  
is the main situation 
that requires a statistical adjustment.  In fact, there are several situations that may confuse the 
researcher by leading to unnecessary or improper adjustments for a third variable.  One case is 
encountered when variable $u$ has an intermediate effect along the pathway from $z$ to $y$ (Figure 
\ref{Fig1} panel (\emph{b})). Controlling for $u$ may be advised if we are interested in the direct 
effect of $z$ on $y$ (i.e.,
that part of effect not mediate through $u$), although it could be quite 
problematic \citep{Freed:99, green:99, cox:wer:04}. On the other hand, adjustment is useless if we 
are interested in the total effect of $z$ on $y$. As outlined in Section \ref{sec:2}, this is the
case with some variables in our study, such as IUGR. 

Another situation takes place  when both $z$ and $y$ have 
a common effect $u$ (Figure \ref{Fig1} panel (\emph{c})): as outlined by \cite{green:99}, in this case 
adjustment for $u$ ``is not only unnecessary but irremediably harmful", as it would result in the 
so-called adjustment-induced bias. 

A final naive situation is illustrated in Figure \ref{Fig1} panel 
(\emph{d}), where two causes of $y$ act independently one another (as shown through the missing edge 
between $z$ and $u$). Therefore, if our interest is limited to the causal effect of $z$ on $y$, 
ignoring $u$ has no consequences.
\subsection{Preliminaries on structural equation models and extensions}  
As mentioned in Section \ref{sec:intro}, 
an useful statistical instrument to control for confounding bias is represented by SEMs 
\citep{wright:21, gold:72, duncan:75, bol:rabe:skr:08}.  
As discussed by \cite{pearl:98, pearl:2009, pearl:2011} and shown in detail by \cite{cox:wer:04}, the 
partial regression coefficients of a SEM 
can be appropriately interpreted in terms of causal effects on the response variable, given that 
\emph{all} the relevant background variables have been included in the model. This point represents 
one of the major difficulties for the causal analysis. Indeed, as outlined by \cite{muth:89}, 
after having controlled for the observed covariates, the residual unexplained heterogeneity in the 
sample may be still substantial.
Two main solutions have been proposed in the literature to treat this 
problem: (i) one is based on the introduction of a continuous latent effect that assumes different 
parameters at individual level \citep{ans:jed:jag:00} and (ii) the other one, which 
is of interest in our 
contribution, is based on the assumption that   the unobserved heterogeneity may be captured by a 
limited number of (unobserved) groups or classes of individuals.
This latter approach is known as 
finite mixture SEM and it was
introduced independently by \cite{jed:jag:des:1997}, \cite{dol:maas:98}, 
and \cite{arm:stein:witt:99}. 
See also \cite{ver:mag:05} for a clear illustration of different aspects, such as model 
specification without and with covariates, estimation, and model selection, and see
\cite{muth:2002} for a wide overview and classification of different types of SEM.  
More precisely, in a finite mixture SEM  a 
different SEM may be specified for each mixture component, so that  different components are allowed to have 
different parameter values and even different model types. 
In particular, we introduce in each structural equation 
a discrete latent variable $u$, the distribution of which is based on
$K$ support points with specific mass probabilities. In this way, $u$ represents a common cause 
for all the 
responses. Moreover, the model we propose is configured as a special case of finite mixture SEM.  Indeed, we assume that the $K$ latent classes differ one another for different 
intercepts, while the functional form of 
each regression equation and the values of structural coefficients are assumed to be constant among 
the classes. 

With respect 
to standard SEMs that are based on continuous latent variables, the extension to the finite mixture
approach presents some advantages. Firstly,  each mixture component identifies homogeneous classes of 
individuals that have very similar
latent characteristics, so that, in a decisional context, individuals in the same latent class will receive 
the same treatment  \citep{laza:50, laza:henr:68, good:74}. 
Moreover, this assumption allows to estimate the SEM in a semi-parametric way, namely
without formulating any  parametric 
assumption on the latent variable distribution.

Another useful extension of the standard SEM  approach, 
that is considered in the present papers, 
derives from the observation that, in their original formulation, SEMs  are based on 
continuous observed variables, so accommodating only a few types of application. 
A more general framework is obtained by adopting a generalized formulation 
\citep{skr:rabe:04, skr:rabe:05, bol:rabe:skr:08}, 
which allows to take into account mixed types of response, that is 
both continuous and ordinal or binary observed responses, in the same set of structural 
equations. In this regards, since among the putative causes we have categorical variables
for the educational level (with categories: 1 for middle school or less,
2 for high school, and 3 for degree or above) and the marital status (with categories:
0 for married, 1 for not married), 
we introduce a latent continuous variable $z_{il}^*$ underlying each 
observable variable $z_{il}$. In particular, we assume that 
\begin{equation*}
z_{il} = G_l(z_{il}^*),
\end{equation*}
where $G_l(\cdot)$ is a function which may depend on specific parameters according to the different nature of $z_{il}$. We consider the following cases: 
\begin{itemize}
\item when the observed response is of a continuous type, an identity function is adopted,
that is $G_l(z_{il}^*) = z_{il}^*$;
\item when the observed response is binary (i.e., $z_{il} = 0,1$), then
\begin{equation}
G_l(z_{il}^*) = I\{z_{il}^* > 0\},\label{eq:binary_link}
\end{equation}
where $I\{\cdot\}$ is the indicator function 
assuming value 1 if its argument is true and zero otherwise;
\item when the observed response is ordinal with categories $j = 1, \ldots, J_l$, 
we introduce a set of cut-points $\tau_{l1} \geq \ldots \geq \tau_{l,J_l-1}$ and we define
\begin{equation}
G_l(z_{il}^*) = \left\{
\begin{array}{cc}
1 & z_{il}^*\leq-\tau_{l1},\\
2 & -\tau_{l1}<z_{il}^*\leq-\tau_{l2},\\
\vdots & \vdots \\
J & z_{il}^*>-\tau_{l,J_l-1}.
\end{array}
\right.\label{eq:G}
\end{equation}
\end{itemize}
\subsection{The proposed finite mixture SEM} 
In the following, coherently with the theoretical 
model illustrated in Section \ref{sec:2} and in accordance with the notation 
previously introduced, the assumed causal relationships among the considered variables are expressed 
through a SEM composed by $3$ equations for each of the $n$ singleton delivers in the dataset.
Two of these equations are referred to the causes educational level ($z_{i1}$) 
and marital status  ($z_{i2}$), whereas the third one refers to the two correlated birth variable
outcomes: gestational age ($y_{i1}$) and birthweight
($y_{i2}$). Moreover, the vector $\b x_i$ is composed by observations on 
the variables age and squared
age (both centered with respect to the mean value), and on dummies referred to citizenship 
(Italian is the reference category). 

For every deliver $i$, $i=1,\ldots,n$, the generalized linear structural equations are as follows:
\begin{itemize}
\item \underline{Equation 1 (educational level)}: 
we assume that $z_{i1}=G_1(z_{i1}^*)$, with $G_1$ defined as in (\ref{eq:G}) and
\begin{equation}\label{eq:education1}
z_{i1}^*  = \mu_1 + \alpha_{i1} + \b x_{i}\tr\bbeta_1 + \varepsilon_{i1},
\end{equation}
where $\mu_1+\al_{i1}$ is a specific intercept for subject $i$, $\b\be_1$ is
a vector of regression coefficients for the covariates in $\b x_i$, and $\varepsilon_{i1}$
is a random error term with logistic distribution;
\item \underline{Equation 2 (marital status)}: we assume that $z_{i2}=G_2(z_{i2}^*)$, 
with $G_2$ defined as in (\ref{eq:binary_link}) and
\begin{equation}\label{eq:marital1}
z_{i2}^* = \mu_2 + \alpha_{i2} + \b x_{i}\tr\bbeta_2   + z_{i1}\tr\ga + \varepsilon_{i2}, 
\end{equation}
where $\mu_2 + \alpha_{i2}$ is the subject specific intercept, $\b\be_2$ and $\ga$
are regression coefficients, and $\varepsilon_{i2}$ is an error term with logistic distribution,
which is independent of $\varepsilon_{i1}$;
\item \underline{Equation 3 (gestational age, birthweight)}: in this case the
observable variables, that we collect in the vector $\b y_{i} = (y_{i1}, y_{i2})\tr$,
are continuous and then we directly assume that
\begin{equation}\label{eq:bivar}
\b y_{i}   =  \bnu + \b\de_i + \b\Phi\b x_{i}+\b\Psi\b z_i + \b\eta_i,
\end{equation}
where $\bnu = (\nu_{1}, \nu_{2})\tr$,  $\b\de_i = (\de_{i1}, \de_{i2})\tr$,  
$\b\Phi = (\b\phi_{1},\b\phi_{2})\tr$, $\b\Psi = (\b\psi_1,\b\psi_2)\tr$,
and $\boeta = (\eta_{1}, \eta_{2})\tr$; the last is a vector of error terms,
which is assumed to follow a bivariate Normal distribution centered at $\b 0$ and
with variance-covariance matrix $\b\Sigma$ and to be independent of the previous
error terms. Here, we have two subject-specific intercepts,
that is $\nu_1+\de_{i1}$ for the gestational age and $\nu_2+\de_{i2}$ for the 
birthweight. Accordingly, we have specific regression coefficients which are
collected in $\b\phi_1$ and $\b\psi_1$ for the first response variable and in $\b\phi_2$
and $\b\psi_2$ for the second one.
\end{itemize}

Commenting the above equations, we note that the parameters of most interest in
the present causal analysis are the regression coefficients in $\b\Psi$. Moreover,
concerning the individual-specific parameters $\al_{i1}$, $\al_{i2}$, and $\b\de_i$,
we clarify that, due to the latent class assumptions described in the previous
section, these parameters have a discrete distribution with $K$ support points
and corresponding probabilities (or weights). In particular, for the $k$-th class,
with $k=1,\ldots,K$,
the support points for $\al_{i1}$ and $\al_{i2}$ are denoted by $\xi_{k1}$ and
$\xi_{k2}$ respectively, whereas the vector of support points for $\b\de_i$ is denoted by
$\b\zeta_k$; 
the corresponding weight is denoted by $\pi_k$. Support points and class weights are
estimated on the basis of the data, together with the other parameters involved
in the previous structural equations.

In order to make the
model identifiable, the support points are suitably constrained by fixing their 
(weighted) mean at 0, so that $\mu_1$, $\mu_2$, and $\b\nu$ have the role average of intercepts. 
Note that, in order to make the model identifiable, we also constraint the
first cutpoint involved in the second equation ($\tau_{21}$) to be equal to 0.

Finally, about the error terms involved in the three equations, we note that 
$\varepsilon_1$ and $\varepsilon_2$ are assumed to have a logistic distribution.
Then, due to the specified $G_l(\cdot)$ functions, a global (or cumulative proportional odds) logit 
parameterization, used in the proportional-odds model of \cite{mccu:80},
results for the conditional distribution of $z_{i1}$, 
whereas a standard logit parametrization results for $z_{i2}$; see also \cite{agre:02}.
In fact, we have that
\begin{equation}\label{eq:education2}
\log \frac{p(z_{i1}\geq j|\al_{i1},\b x_i)}{p(z_{i1} < j|\al_{i1},\b x_i)} =
\mu_1+\tau_{1,j-1}+\alpha_{i1}+\b x_i\tr\bbeta_1,\quad j = 2,3,
\end{equation}
and
\begin{equation}\label{eq:marital2}
\log\frac{p(z_{i2} = 1|\al_{i2},\b x_i,z_{i1})}{p(z_{i2} = 0|\al_{i2},\b x_i,z_{i1})} =  
\mu_2 + \alpha_{i2} + \b x_{i}' \bbeta_2+z_{i1}\tr\b\ga.
\end{equation}
Note that, we could also assume that both $\varepsilon_1$ and $\varepsilon_2$ have a Normal
distribution, resulting in a parametrization based on probits and ordered probits,
 but this
would have small impact on the model specification, while making the estimation more
complex. On the other hand, given the nature of the variables in $\b y_i$, it is obvious
to assume a bivariate Normal distribution in the third equation for $\b y_i$. Note,
in particular, that this distribution depends on the variance-covariance matrix
\[
\b\Sigma = \left[
\begin{array}{cc}
\sigma_1^2 &\sigma_{12}\\
\sigma_{21} &  \sigma_2^2 \\
\end{array}
\right],
\]
where $\sigma_1^2$ is the variance of $y_1$, $\sigma_2^2$ is the variance of $y_2$, and 
$\sigma_{12} = \sigma_{21}$ is the covariance between $y_1$ and $y_2$. All are free parameters
and then we allow a free correlation, which is measured by the index
\[
\rho = \frac{\si_{12}}{\sqrt{\si_1^2\si_2^2}},
\]
between the two variables, even given the observable and the unobservable variables.
\subsection{Model estimation}
We perform estimation of the parameters of the model previously introduced 
by the maximum likelihood method. This requires to derive
the {\em joint distribution} of ($z_{i1}, z_{i2}, \b y_{i}$), that is
the conditional distribution of these variables given $\b x_i$, once the individual-specific
parameters $\al_{i1}$, $\al_{i2}$, and $\b\de_i$ have been integrated out. This
distribution may be expressed as a finite mixture, that is
\[
f(z_{i1}, z_{i2}, \b y_i | \b x_i) = \sum_{k=1}^K \pi_k p(z_{i1}|\al_{i1}=\xi_{k1},\b x_i)
p(z_{i2}|\al_{i2}
=\xi_{k2},\b x_i, z_{i1})f(\b y_i|\b\de_i=\b\zeta_k,\b x_i, \b z_i),
\] 
where the probabilities mass function for $z_{i1}$ and $z_{i2}$ are defined through
(\ref{eq:education2}) and (\ref{eq:marital2}), whereas 
$f(\b y_i|\b\de_i=\b\zeta_k,\b x_i,\b z_i)$ is the
density function, computed at $\b y_i$, of a bivariate normal distribution
with mean $\bnu + \b\de_i + \b\Phi\b x_{i}+\b\Psi\b z_i$
and variance-covariance matrix $\b\Sigma$; see equation (\ref{eq:bivar}).  

Under the assumption that the sample units are independent each other, 
the log-likelihood of the model to be maximized for the estimation is
\[
\ell(\b\theta) = \sum_{i=1}^n \log f(z_{i1}, z_{i2}, \b y_i | \b x_i),
\]
with $\b\theta$ denoting the vector of all model parameters. 
Maximization of $\ell(\b\theta)$ may be efficiently performed through an
Expectation-Maximization (EM) algorithm 
\citep{demp:lair:rubi:77}. In the following, we sketch this algorithm, referring to 
more specialized papers on the topic; see, for instance, \cite{bart:forc:06} and
the references therein. Moreover, as already mentioned, for the present application
we implemented the algorithm in a set of {\tt R} functions that we make available
to the reader upon request.

The EM algorithm is based on the so-called {\em complete data log-likelihood},
which could be computed if we knew the latent class to which every sample
unit belongs. This function may be expressed as
\begin{eqnarray}
\ell^*(\b\theta) &=& \sum_{k=1}^K\sum_{i=1}^n w_{ik}\log[p(z_{i1}|\al_{i1}=\xi_{k1},\b x_i)
p(z_{i2}|\al_{i2}=\xi_{k2}, \b x_i, z_{i1})f(\b y_i|\b\de_i=\b\zeta_k,\b x_i, \b z_i)]+\nonumber\\
&+&\sum_{k=1}^K\sum_{i=1}^nw_{ki}\log\pi_k,\label{eq:comp_lk}
\end{eqnarray}
where $w_{ik}$ is a dummy variable equal to 1 if unit $i$ belongs to subject $k$
and to 0 otherwise. Based of this function, the EM algorithm alternates 
the following two steps until convergence in $\ell(\b\theta)$:
\begin{itemize}
\item \underline{step E}: compute the conditional expected value of $\ell^*(\b\theta)$
given the observed data and the current value of the parameters in $\b\th$;
this is equivalent to substituting every dummy variable $w_{ik}$ in (\ref{eq:comp_lk})
by the corresponding conditional expected value
\[
\hat{w}_{ik}=\frac{\pi_kp(z_{i1}|\al_{i1}=\xi_{k1},\b x_i)
p(z_{i2}|\al_{i2}=\xi_{k2},\b x_i, z_{i1})f(\b y_i|\b\de_i=\b\zeta_k,\b x_i, \b z_i)}
{f(z_{i1}, z_{i2}, \b y_i | \b x_i)};
\]
\item \underline{step M}: maximize the expected value of $\ell^*(\b\th)$ obtained above
with respect to the model parameters; to update the class weights we have an explicit
solution given by
\[
\pi_k=\frac{\sum_{i=1}^n\hat{w}_{ik}}{n},\quad k=1,\ldots,K.
\]
Moreover, for the parameters involved in (\ref{eq:education1}) and (\ref{eq:marital1}) we use
simple iterative algorithms that are currently used to maximize the weighted log-likelihood of a 
proportional odds model \citep{mccu:80}, whereas the parameters in equation (\ref{eq:bivar}) 
are updated by solving a weight least square problem.
\end{itemize}
 
The value of $\b\th$ at convergence of the EM algorithm is taken as the
maximum likelihood estimate of this parameter vector, denoted by $\hat{\b\theta}$.
To treat the well-known problem of multi-modality of 
likelihood characterizing finite mixture and latent variable models, 
we suggest to initialize the estimation
algorithm by both deterministic and random starting values. Finally, standard
errors for the parameter estimates are obtained by inversion of the observed
information matrix, which is numerically obtained from the score function that,
in turn, is obtained by exploying a result due to \cite{oake:99}.
\section{Results}
In the following, we illustrate the results obtained through the finite mixture SEM presented in the 
previous section and 
applied to the dataset about the 9,005 newborns collected in the Region of Umbria;
see Section \ref{sec:data} for a description of the dataset.
Firstly, we give the results about the selection process of the optimal number of latent classes 
(Table \ref{tab:BIC}). Secondly, the estimated regression coefficients are reported separately for 
each structural equation in Tables \ref{tab:education}, \ref{tab:marital}, and \ref{tab:biv}.
Finally, the adopted latent structure is described in Table \ref{tab:latent}, 
which shows the estimated support points for each latent class and the corresponding weights.

First of all, in applying the proposed finite mixture approach, the choice
of the optimal number of latent classes $K$ is of crucial importance. 
For this aim, several studies \citep{roe:97, das:98, mcla:peel:00}
conclude that the Bayesian Information Criterion \citep[BIC;][]{sch:78} 
present an adequate performance for choosing $K$. We remind that this 
criterion is based on penalizing the maximum value of the log-likelihood $\hat\ell$ by a term depending on the 
number of free parameters ($\# \mathrm{par}$) and on the sample size ($n$):
\[
\mathrm{BIC} = -2\hat\ell + \log(n)\# \mathrm{par}.
\]
In practice, we fit the adopted SEM 
with increasing  $K$ values, relying the choice of optimal $K$ on
the value just before the first increasing of the BIC index. 
On the basis of results shown in Table \ref{tab:BIC}, we obtain the minimum BIC value in 
correspondence of $K = 3$ latent classes. 

\begin{table}[!h]
\begin{center}
\begin{tabular}{cccc}
  \hline
$K$ & $\hat\ell$ & $\#$par & BIC \\ 
  \hline
1 & -35700.768 & 32 & 71692.914 \\ 
  2 & -34536.422 & 37 & 69409.750 \\ 
  3 & -34488.589 & 42 & 69359.610 \\ 
  4 & -34467.548 & 47 & 69363.055 \\ 
   \hline
\end{tabular}
\caption{\em Results from the preliminary fitting: number of mixture components ($K$), maximum log-likelihood ($\hat\ell$), number of parameters ($\#$par), and BIC index}\label{tab:BIC}
\end{center}
\end{table}

We now consider the estimates for the structural equations 
(\ref{eq:education1}) and (\ref{eq:marital1})
about education and marital status, respectively.
In particular, we observe that the educational level significantly increases with 
the age of the woman and it is lower 
for foreigners with respect to Italians (Table \ref{tab:education}). Effects of age and citizenship 
are also highly significant with reference to the marital status (Table \ref{tab:marital}): the 
probability to be married increases with age and it is higher for foreign women. Moreover, a causal 
effect of educational level on marital status is detected after controlling for the latent classes: 
higher the educational level, higher the probability to be married. 

\begin{table}[ht]
\begin{center}
\begin{tabular}{lllrrrr}
  \hline
\multicolumn1c{covariate} & \multicolumn1c{category}       & \multicolumn1c{est.} & \multicolumn1c{s.e.}  & \multicolumn1c{$t$ stat.}   & \multicolumn1c{$p$-value} \\ 
  \hline
 intercept ($\mu_1$) & \multicolumn1c{--} & \;2.053 & 0.039 & 52.285 & 0.000 \\ 
1st cutpoint ($\tau_1$) & \multicolumn1c{--} & \;0.000 & \multicolumn1c{--}  & \multicolumn1c{--}  & \multicolumn1c{--}  \\ 
 2st cutpoint ($\tau_2$) &\multicolumn1c{--} & -2.695 & 0.031 & -20.780 & 0.000 \\ 
  age & \multicolumn1c{--}  & \;0.103 & 0.004 & 23.405 & 0.000 \\ 
  age$^2$ &  \multicolumn1c{--} & -0.009 & 0.001 & -14.587 & 0.000 \\ 
       citizenship & Italian & \;0.000 & \multicolumn1c{--} & \multicolumn1c{--} & \multicolumn1c{--} \\
  citizenship & east-Europa & -0.806 & 0.069 & -11.712 & 0.000 \\ 
  citizenship & other citizenship & -1.100 & 0.086 & -12.780 & 0.000 \\ 
   \hline
\end{tabular}
\caption{\em Regression results for educational level}
\label{tab:education}
\end{center}
\end{table}

\begin{table}[ht]
\begin{center}
\begin{tabular}{llrrrr}
  \hline
 \multicolumn1c{covariate} & \multicolumn1c{category}       & \multicolumn1c{est.} & \multicolumn1c{s.e.}  & \multicolumn1c{$t$ stat.}   & \multicolumn1c{$p$-value} \\ 
  \hline
intercept ($\mu_2$) & \multicolumn1c{--}  & -0.763 & 0.065 & -11.313 & 0.000 \\ 
  age & \multicolumn1c{--}  & -0.027 & 0.005 & -5.487 & 0.000 \\ 
  age$^2$ & \multicolumn1c{--}  & 0.008 & 0.001 & 12.381 & 0.000 \\ 
   citizenship & Italian & 0.000 & \multicolumn1c{--} & \multicolumn1c{--} & \multicolumn1c{--} \\
  citizenship & east-Europa & -0.679 & 0.082 & -8.264 & 0.000 \\ 
  citizenship & other citizenship & -0.677 & 0.101 & -6.701 & 0.000 \\ 
         education & middle school or less & 0.000 & \multicolumn1c{--} & \multicolumn1c{--} & \multicolumn1c{--} \\
  education & high school & -0.152 & 0.064 & -2.375 & 0.018 \\ 
  education & degree or above & -0.468 & 0.076 & -6.123 & 0.000 \\ 
   \hline
\end{tabular}
\caption{\em Regression results for marital status}
\label{tab:marital}
\end{center}
\end{table}

We now analyze the results shown in Table \ref{tab:biv} about the outcome 
variables, according to the formulation of equation (\ref{eq:bivar}).
Both gestational age and birthweight decrease as mother's age increases, 
while the association with citizenship is something different. Women from east-Europa 
deliver significantly before Italians, but their newborns have a higher weight.
On the other hand, women from other countries present differences on both response 
variables with respects to Italian women.

\begin{table}[!h]
\begin{center}
\begin{tabular}{lllrrrr}
  \hline
 \multicolumn1c{response var.} &  \multicolumn1c{covariate} & \multicolumn1c{category}       & \multicolumn1c{est.} & \multicolumn1c{s.e.}  & \multicolumn1c{$t$ stat.}   & \multicolumn1c{$p$-value} \\ 
  \hline
Gestational age & intercept ($\nu_{1}$) & \multicolumn1c{--} & 39.346 & 0.044 & 905.935 & 0.000 \\ 
  & age & \multicolumn1c{--}   & -0.015 & 0.003 & -4.789 & 0.000 \\ 
 & age$^2$ & \multicolumn1c{--}   & -0.001 & 0.000 & -2.544 & 0.011 \\ 
   & citizenship & Italian & 0.000 & \multicolumn1c{--} & \multicolumn1c{--} & \multicolumn1c{--} \\
 & citizenship  & east-Europa & -0.194 & 0.049 & -3.942 & 0.000 \\ 
 & citizenship & other citizenship & -0.112 & 0.060 & -1.855 & 0.064 \\ 
   &     education & middle school or less & 0.000 & \multicolumn1c{--} & \multicolumn1c{--} & \multicolumn1c{--} \\
 & education & high school & 0.025 & 0.042 & 0.608 & 0.543 \\ 
 & education & degree or above & 0.029 & 0.049 & 0.600 & 0.548 \\ 
    &     marital & married & 0.000 & \multicolumn1c{--} & \multicolumn1c{--} & \multicolumn1c{--} \\
 & marital & not married & 0.025 & 0.033 & 0.749 & 0.454 \\ 
  \hline
Birthweight & intercept ($\nu_{2}$) & \multicolumn1c{--}  & 3.238 & 0.017 & 195.392 & 0.000 \\  
 & age & \multicolumn1c{--}  & -0.004 & 0.001 & -3.863 &0.000 \\ 
 & age$^2$ &   \multicolumn1c{--}  & -0.000 & 0.000 & -1.708 & 0.088 \\ 
   & citizenship & Italian & 0.000 & \multicolumn1c{--} & \multicolumn1c{--} & \multicolumn1c{--} \\
 & citizenship & east-Europa & 0.041 & 0.016 & 2.653 & 0.008\\ 
 & citizenship & other citizenship & -0.031 & 0.019 & -1.608 & 0.108 \\ 
    &     education & middle school or less & 0.000 & \multicolumn1c{--} & \multicolumn1c{--} & \multicolumn1c{--} \\
 & education & high school & 0.023 & 0.014 & 1.674 & 0.094 \\ 
 & education & degree or above & 0.043 & 0.017 & 2.462 & 0.014 \\ 
     &     marital & married & 0.000 & \multicolumn1c{--} & \multicolumn1c{--} & \multicolumn1c{--} \\
 & marital & not married & 0.011 & 0.012 & 0.904 & 0.366 \\ 
   \hline
 &  \multicolumn2l{variance of gestational age ($\sigma_1^2$)} &  1.776 & \\ 
  & \multicolumn2l{variance of birthweight ($\sigma_2^2$)} &  0.171 & \\ 
 & \multicolumn2l{covariance ($\sigma_{12}$)} &  0.248  & \\ 
 & \multicolumn2l{correlation ($\rho_{12}$)} & 0.450 & \\ 
\hline
\end{tabular}
\caption{\em Regression results for the gestational age and the birthweight}
\label{tab:biv}
\end{center}
\end{table}

By comparing Table \ref{tab:biv} with Tables \ref{tab:reg_gest} and \ref{tab:reg_weight}, it is 
possible to draw some conclusions about the causal effects of educational level and marital status on 
the birth outcomes. About the marital status, the analysis confirms the absence of any causal 
effect. With reference to the educational level, the increase of $p$-values denote the presence of a 
confounding effect. However, even after controlling for a latent common cause, a significative effect 
persists on the birthweight: \emph{a higher educational level causes a higher birthweight}. 
Note that the causal effect is well defined for 
the degree or above modality, with a $p$-value 
equal to $0.014$, while this is not longer true for mothers with a high school.
Moreover, we observe that correlation between outcomes 
is not enough large ($\rho_{12}=0.45$). Given that gestational age and birthweight are conditioned to 
covariates and latent variables, this is compatible with causes affecting
in a different way the two outcomes.

We now analyze the results concerning the latent structure of the model 
(Table \ref{tab:latent}). As mentioned above, $K=3$ different latent classes are detected. Note 
that the estimated $p$-values refer to the comparisons between classes 2 or 3 versus class 1. The 
most representative class is the first one, to which 
corresponds a weight equal to  ${\pi}_1 = 0.931$, whereas the remaining part of women results 
assigned to the third class with ${\pi}_3 = 0.041$ and to the second class with ${\pi}_2 = 
0.028$.

\begin{table}[!h]
\begin{center}
\begin{tabular}{llrrr}
  \hline
\multicolumn1c{} & \multicolumn1c{$k=1$} & \multicolumn1c{$k=2$}       & \multicolumn1c{$k=3$}  \\ 
  \hline
  education ($\xi_{k1}$) &   0.005 &     -0.165 (0.234) &    -0.005  (0.964) \\
  marital status ($\xi_{k2}$)  & 0.026 & 0.289 (0.081) &  -0.794 (0.021) \\
  gestational age ($\zeta_{k1}$)  &   0.178 &   -6.086 (0.000)  &  0.123 (0.671) \\
    birthweight ($\zeta_{k2}$) &    0.005 &   -1.245 (0.000) &  0.728 (0.000) \\
    \hline
       class weight ($\pi_k$) & 0.931    &     \multicolumn1c{0.028}         & \multicolumn1c{0.041}   \\
 \hline  
  \end{tabular}
\caption{\em Class weights and support points estimates (p-values in parenthesis are referred to
the comparison between the second and third class with the first class)}
\label{tab:latent}
\end{center}
\end{table}

With a weight greater than 0.90, women belonging to class 1 represent the main part of 
the population, so that no particular difference results with respect to the average 
values of the entire population, as shown by the estimated support points which very close to zero. 

Differently, women from class 2 result to be well characterized. With a support point equal 
to 0.289, to which correspond an odds equal to $\exp(0.289) =1.335$, 
women in class 2 present a significant
higher propensity (at $10\%$ level) to be not married with respect to women in
the first class. On average they give birth  
6.1 weeks before and their infants weigh 1.245 kg less.
No significant difference results about the educational level. Finally, 
women in class 3 have a higher tendency to be married (odds equal to $\exp(-0.794)=0.452$
for not being married) and the birthweight of her infants is  
significantly higher ($+0.728$ kg with respect to the first class). On the other hand,  no 
significant difference results with respect to educational level and gestational age.

Our estimates allow us to assess potential heterogeneity in infant health outcomes of educational 
attainments across mothers. We focus our attention on birthweight because, differently from 
recent findings of this literature, education does not find significant estimates for gestational 
age. The estimates suggest that higher education increases birthweight by a significant, although 
quantitatively not large, amount: a graduate mother increases, on average, of 43 gr the newborn 
weight with respect to mothers with a basic educational level.
This impact on birthweight is more 
limited for mothers that only achieve a high school level; it mostly cast some doubt if 
the estimated parameters may or may not impact on child health, because the significance is between 
5\% and 10\% level. Three facts stand out from our results, as we explain in the following.

First, the finite mixture SEM estimates for the effect of
education on birthweight are lower than 
standard regression ones (and with higher $p$-values), a finding 
common to studies similar to ours \citep{abrevaya:08}, 
and in line with the a priori that upward 
bias in regression estimates are driven by the correlation between mother's education and unobserved 
variables.

Second, the finite mixture SEM suggests a significant and positive effect of education on the 
probability to be married. Specifically, non-completing high school or degree would lower the mother 
probability to be married of about 14\% and 37\%, respectively.
While the range of variation of these estimates may reflect the specificity of our sample, we note 
that in most cases they are against the estimates reported in this specific literature. See, for 
example, \cite{breierova:04} for a study in Indonesia and \cite{Lefgren:06} for a study in the US. 
As known, based on catholic background of marriage, the postponement of mother with higher education
in Italy does not influence their propensity towards traditional cohabitation.

Third, the estimated effect of the latent classes stems qualitatively almost exclusively from 
differences between married and not married mothers or, alternatively, within the married group.

Different arguments may help to explain the first result. The prevalent interpretation is that 
the woman's educational level may be related with specific unobservable variables, such
as the ability to properly manage the pregnancy so as to improve the health level of the newborn. 
This implies that an upward bias emerges in the 
regression estimates, a result consistent with the findings of Table 4. 
On the other hand, unlike \cite{abrevaya:08}, we find a greter
effect on birthweight 
for mothers with the highest level of education (with respect to women with a high school degree), 
a result qualitatively in accordance with 
\cite{currie:03}\footnote{However, the recent literature that use laws affecting the compulsory 
schooling of high school educated
mothers has not shown a positive impact on birthweight \citep{Lindeboom:09,McCrary:11}.}. 
Note that, 
if birthweight responds to educational level, conditionally
on to be married or not, unobservable factors linked 
with her husband may further affect the results, although latent classes have been identified. We 
will return to this discussion below, when we attempt to identify a latent class by a positive 
assortative mating within married mothers. 
 
Despite a growing literature, little is known about to the causal effect of the
woman's educational level on her marital status. Our estimates indicate that the 
education level has a positive impact on the marriage probability, especially
with an educated potential partner. This is clearly not in line 
with the results of \cite{breierova:04}, in which an increase of
education does not lead to a significant impact on the probability
of a woman being currently married.
Although there are several respects in which the increase in the marriage chance
is consistent with a higher level of education, we can
conjecture that our data are in line with the hypothesis that this effect is mitigated by the 
specificity of the Italian labor market. The job participation rate of the mothers in our sample
is close to that of the national mean of about 45\%, an index distant of more 
than 10 percentage points from the average of the EU countries and far from the aims of 
European Pact for gender equality\footnote{EU Council conclusions 7370/11 of 08/03/2011.}. 
In addition, this is heterogeneous across levels of education.
In fact, if we refer to mothers with at least 
a degree before the first pregnancy, the Italian participation rate is even much lower.
In contrast to the classical predictions in terms of incentive marriage \citep{lam:88}, this low 
participation rate of women may induce potential gains from household 
specialization. 
The result reinforces the known evidence that more educated mothers use the status to be married to 
postpone the job search or to avoid to select jobs that are not in accordance 
to their individual expectations, a finding not 
in contrast with the stylized fact that, over the life cycle, 
more educated mothers have on average a higher participation rate.  

By acquiring the highest educational level, a mother can affect the identity of his future
husband. However, the identification of the effect of education on the probability 
to be married and outcomes at birth is complicated by the
endogeneity involved in the partner's educational level, 
that may affect mothers' choices and outcomes; that is, marriages
may well be determined by factors such as social background and geographical location.
These factors are also correlated with education, and could lead the observed correlation in 
spouses' education to be partly or entirely spurious. 
We have recognized above that the last latent class includes women
with a higher chance to be married.
The large and positive impact on birthweight of the third class may 
lead to assume a married mothers' group with a positive assortative mating. 
Here, we indirectly investigate the latter effect by including in the finite mixture class 
SEM the father's educational level as a potential confounder of the birthweight outcome and evaluate 
the sensitivity with respect to the main relationship between education and birthweight.
Tables \ref{tab:biv_titpa} and \ref{tab:latent_titpa} shows the results of these estimates. 
Noticeable, the contribution of education on birth outcome becomes less susceptible to ambiguous 
conclusions. Causal effect of mother's education with at least a degree are still significant and 
large enough to conclude over the goodness of the estimates, while it is even clearer that those with 
high school do not contribute to explain differences in the birthweight. Moreover, our findings 
appear first to support the idea that the mothers in class 3 have a positive assortative mating with  
a large effect on infant outcomes (compare $\hat\gamma_{32}$ of Table \ref{tab:latent_titpa} with its 
analogous of Table \ref{tab:latent}), and to emphasize the robustness of our analysis, we note that 
all other estimated coefficients are close to the model presented in table \ref{tab:biv}.

\begin{table}[!h]
\begin{center}
\begin{tabular}{lllrrrr}
  \hline
 \multicolumn1c{response var.} &  \multicolumn1c{covariate} & \multicolumn1c{category}       & \multicolumn1c{est.} & \multicolumn1c{s.e.}  & \multicolumn1c{$t$ stat.}   & \multicolumn1c{$p$-value} \\ 
  \hline
Gestational age &intercept ($\nu_1$) & \multicolumn1c{--} & 39.473 &	0.109	&	362.26	&	0.000	\\ 
    &     education & middle school or less & 0.000 & \multicolumn1c{--} & \multicolumn1c{--} & \multicolumn1c{--} \\
 & education & high school & 0.012	&	0.043	&	0.27	&	0.787	\\ 
 & education & degree or above & 0.021	&	0.053	&	0.40	&	0.690	\\ 
  & marital & married & 0.000	&	 \multicolumn1c{--} 	&	 \multicolumn1c{--} 	&	 \multicolumn1c{--} 	\\ 
 & marital & not married & 0.013	&	0.034	&	0.39	&	0.699	\\ 
  \hline
Birthweight & intercept ($\nu_2$) & \multicolumn1c{--}  & 3.263	&	0.115	&	28.54	&	0.000	\\
    &     education & middle school or less & 0.000 & \multicolumn1c{--} & \multicolumn1c{--} & \multicolumn1c{--} \\
 & education & high school & 0.015	&	0.014	&	1.11	&	0.268	\\
 & education & degree or above & 0.032	&	0.017	&	1.93	&	0.053	\\
   & marital & married & 0.000	&	 \multicolumn1c{--} 	&	 \multicolumn1c{--} 	&	 \multicolumn1c{--} 	\\ 
 & marital & not married & -0.007	&	0.011	&	-0.69	&	0.489	\\ 
\hline
\end{tabular}
\caption{\em Regression results for the gestational age and the birthweight controlled for father's educational level, age and citizenship}
\label{tab:biv_titpa}
\end{center}
\end{table}

\begin{table}[!h]
\begin{center}
\begin{tabular}{llrrr}
  \hline
\multicolumn1c{} & \multicolumn1c{k=1} & \multicolumn1c{k=2}       & \multicolumn1c{k=3} \\ 
  \hline
  gestational age ($\zeta_{k1}$) & 0.173  & -5.995 (0.000) &  0.172  (0.993)	\\
   birthweight ($\zeta_{k2}$) &   0.035 &  -1.215  (0.000) & 0.036    (0.993)	\\
 \hline
   \end{tabular}
\caption{\em Support points estimates after controlling for father's educational level (p-values in parenthesis)}
\label{tab:latent_titpa}
\end{center}
\end{table}

\section{Conclusions}

The article presents new evidence on the child health increasing effect of education, using a
finite mixture structural equation model (SEM)
to identify a causal link. In particular, the estimation strategy controlling for the effects of 
marital status, observed and unobserved characteristics of
the mother guarantees that health outcomes are entirely driven by differences in
education. This is made possible by the inclusion
of random parameters in each structural equation, which follow a discrete distribution with support points and weights
estimated on the basis of the dataset. These support points then identify latent classes
of individuals that allow us to adjust for unobserved confounding.

We report empirical findings showing that high education of mothers increases the birthweight whereas
gestational age is not affected. The estimated social saving from birthweight increase, implied by 
our estimates, are substantial if associated to married mothers with positive assortative mating, 
that we identify through the latent classes. 

The existence of a causal birthweight increasing effect of high education has potentially important 
implications for longer-term effort aimed at reducing the level of birthweight. Policies that 
incentive high school, as an investment in human capital, have significant potential to reduce the 
percentage of mothers that deliver children below the minimum threshold weight, by increasing skill 
levels in helping care and behaviors consistent with a healthy pregnancy.
At the very least, our result confirm that improving education among young mothers should be viewed 
as a key policy to reduce costs of unhealthy child outcome, a finding that appear to be emphasized at 
the ``family level'' by the presence of a more educated husband.

\bibliography{biblio}
\bibliographystyle{apalike}

\end{document}